\documentclass[12pt]{article}
\usepackage{indentfirst}
\usepackage[titletoc]{appendix}
\usepackage{graphicx}
\usepackage[hypcap]{caption}
\usepackage{hyperref}
\usepackage{amsmath, amsthm, amssymb}
\usepackage{array}
\usepackage{booktabs}
\title{THE STORY OF THE NEUTRINO}
\author{G Rajasekaran \\
Institute of Mathematical Sciences, Taramani, Chennai\\
\& Chennai Mathematical Institute, Siruseri, Chennai\\
email : \texttt{graj@imsc.res.in} }

\date{}
\begin{document}

\maketitle

\begin{abstract}
	This is an elementary review of the history and physics of neutrinos.  The story of the discovery of neutrino mass through neutrino oscillations is described in some detail.  Experiments on solar neutrinos and atmospheric neutrinos played an important part.  Recent advances are summarized and future developments are indicated.
\end{abstract}

\pagebreak
% % \begin{abstract}
% % \end{abstract}
\tableofcontents

\pagebreak
\section{Introduction}
In the recent past, two Nobel Prizes were given to Neutrino
Physics. In 2002 Ray Davis of USA and Matoshi Koshiba of
Japan got the Nobel Prize for Physics while last year (2015)
Arthur McDonald of Canada and Takaaki Kajita of Japan got
the Nobel Prize. To understand the importance of neutrino
research it is necessary to go through the story of the neutrino
in some detail.

Starting with Pauli and Fermi, the early history of the neutrino 
is described culminating in its experimental detection 
by Cowan and Reines. Because of its historical
importance the genesis of the solar neutrino problem and
its solution in terms of neutrino oscillation are described
in greater detail. In particular, we trace the story of the
90-year-old thermonuclear hypothesis which states that the
Sun and the stars are powered by thermonuclear fusion reactions
and the attempts to prove this hypothesis experimentally. We go  
through Davis's pioneering experiments to detect the neutrinos 
emitted from these reactions in the Sun and  
describe how the Sudbury Neutrino Observatory in Canada was
finally able to give a direct experimental proof of this
hypothesis in 2002 and how, in the process, a fundamental
discovery i.e. the discovery of neutrino oscillation and neutrino
mass was made.

We next describe the parallel story of cosmic-ray-produced neutrinos
and how their study by SuperKamioka experiment in Japan won the
race by discovering neutrino oscillations in 1998. 

Many other important issues are briefly discussed at the end.

\section{What is a Neutrino?}
Neutrino is an elementary particle like electron. But unlike 
electron which has a negative electric charge, it is neutral.
Also, unlike electrons which are constituents of all atoms,
neutrinos do not exist within atoms. But they are created
through many processes all over the Universe in large numbers
and are flying everywhere at almost the speed of light.
Every second more than $ 10^{12} $ neutrinos are passing 
through our body without affecting us in any way. Since the
probability of neutrinos interacting with matter is
negligible, they simply pass through all matter. Hence it
requires huge detectors and sophisticated instruments to
study them. 

Until some years ago, neutrinos were regarded as massless
particles like photons. But in 1998 neutrinos were discovered  
to have mass. This discovery is expected to lead to
fundamental changes in our knowledge of physics and astronomy.
Many more discoveries about neutrinos are yet to be made.

\section{Early History of Neutrino}

After radioactivity was discovered by Becquerrel in 1897,
many properties of radioactivity were revealed by the
researches of a host of scientists including the famous
ones Marie Curie and Ernest Rutherford. Among those,
the so-called beta radioactivity turned out to be a puzzle.
The electrons that came out in the beta activity did not
come out with a single energy unlike the case of alpha
and gamma activity where the alpha particle or the gamma
photon emitted by a particular nucleus came with a single
energy. The beta electrons had a continuous spectrum of
energies. This seemed to contradict the principle of
conservation of energy which is a cornerstone of Physics.
Wolfgang Pauli in 1930 suggested a way to resolve this
puzzle. If another unseen particle was emitted along with
the electron, it could take away part of the energy and
thus the principle of conservation of energy could be
saved. This was Pauli's suggestion. 

Although neutrino was born in the mind of Pauli,
it was Enrico Fermi who made neutrino the basis of 
his famous theory of beta decay in 1932 and
showed how in the beta decay of a nucleus an electron
and a neutrino are simultaneously created \cite{1}. It is this 
that remained as the basic theory of the decays of
all elementary particles for more than 40 years. It was also
Fermi who christened the particle as 'Neutrino'.

In the subsequent decades beta decays of many atomic nuclei 
were experimentally studied. All of them were in beautiful agreement with 
Fermi's theory and hence it was clear to theoretical 
physicists at least that Pauli's neutrinos were indeed emitted
in beta decay. But Cowan and Reines did not agree. If
neutrinos exist, their existence must be experimentally proved,
they said. And they proved it in 1954.

Before we describe their experiment, it is necessary to
explain beta decays of nuclei.

\section{Beta decays and the Cowan-Reines Experiment}

Every atomic nucleus contains Z number of protons and N
number of neutrons. For example, the nucleus of the Hydrogen
atom is a single proton. Helium nucleus contains 2 protons
and 2 neutrons. Uranium nucleus contains 92 protons and
146 neutrons. Many nuclei undergo beta decay spontaneously.
The nucleus (Z,N) which contains Z protons and N neutrons
emits an electron $ (e^{-})$ and an antineutrino $ \bar{\nu}_{e} $
and becomes the nucleus (Z+1,N-1) containing Z+1 protons and
N-1 neutrons. This is shown below.
\begin{align}
(Z,N) &\rightarrow \ (Z+1,N-1) + e^- + \bar{\nu_e}\ (\beta^-\ decay)
\end{align}
\noindent
In the same way, neutron (n) decays and becomes a proton as
shown below.
\begin{align}
n &\rightarrow \ p + e^- + \bar{\nu_e}\ (\beta^-\ decay\ of \ n)
\end{align}
\noindent
If we transfer the antineutrino from the right side of the
eq.(1) to the left side, it will be a neutrino. As shown below, this then denotes the reaction in which a neutrino $\nu_{e} $ collides with a nucleus (Z,N) and the result is another nucleus (Z+1, N-1) and an electron $e^{-}$.
\begin{align}
\nu_e\ +\ (Z,N) &\rightarrow \ (Z+1,N-1) + e^-\ (inverse\ \beta\ decay) 
\end{align}
\noindent
This is sometimes called inverse beta decay and it is through such
reactions experimental physicists detected neutrinos.

In eq.(4) shown below, nucleus (Z,N) emits a positron $e^{+}$ and a neutrino
$\nu_{e}$ and becomes the nucleus (Z-1, N+1).
\begin{align}
(Z,N)&\rightarrow \ (Z-1,N+1)+e^++\nu_e\ (\beta^+\ decay)
\end{align}
\noindent
Here, if we transfer the neutrino to the left side, it will be an antineutrino $\bar{\nu}_{e}$ and we will have a reaction of the antineutrino (shown in eq.(5)).
\begin{align}
\bar{\nu_e}+(Z,N)&\rightarrow \ (Z-1,N+1)+e^+\ (Inverse\ \beta^+\ decay)
\end{align}
\noindent
As an example, an antineutrino and  a proton collide and become a
positron and a neutron (eq.(6)). 
\begin{align}
\bar{\nu_e}+p&\rightarrow \ n+e^+\ (Cowan-Reines\ reaction)
\end{align}
\noindent
It is this reaction that Cowan
and Reines used to prove the real existence of the neutrino 
(actually the antineutrino).

Every nuclear reactor is a copious source of antineutrinos. How?
When nuclei such as Uranium fission in the nuclear reactor, a
variety of radioactive nuclei are produced. Many of them
undergo beta decay and emit antineutrinos. Cowan and Reines used
a hydrogenous material as their detector. Hydrogen nucleus is
a proton. If the antineutrino from the reactor interacts with the
proton, a positron and a neutron are produced, as we already
saw (eq.(6)). Reines and Cowan proved the appearance of
the positron and neutron in their detector placed near the nuclear
reactor. Thus the emission of antineutrinos from the nuclear reactor
was experimentally proved by Cowan and Reines in 1954. Reines 
received the Nobel Prize in 1995. Cowan had passed away before that.

There are two interesting episodes connected to the Cowan-Reines experiment.
In that period (1945-55) many nuclear bomb tests were being conducted.
In the explosion of the nuclear bomb also, Uranium nucleus fissions
and antineutrinos are produced. Cowan and Reines had planned to catch
those antineutrinos, but were prevented from pursuing that dangerous
venture. They then changed their plan and went to the Savannah River
Reactor (USA) to do their experiment and succeeded. Pauli had
apparently sent a cable telegram to the Committee which was to decide
on the sanction of financial support for the Cowan-Reines experiment,
saying that "his particle" cannot be detected by anybody and so asking
the Committee not to support such an experiment. However that telegram
did not reach the Committee in time; support was given and the antineutrino
was caught in the experiment!

\section{Neutrinos from the Sun}

It is the Sun that is giving us light and heat. Without it,
life on Earth is impossible. How does Sun produce its energy
and continue to shine for billions of years?
In the 19th century, the source of the energy in the Sun
and the stars remained a major puzzle in science, which
led to many controversies. Finally, after the
discovery of the atomic nucleus and the tremendous amount of energy locked up
in the nucleus, Eddington in 1920 suggested nuclear energy as
the source of solar and stellar energy. It took
many more years for the development of nuclear physics
to advance to the stage when Bethe,the Master Nuclear 
Physicist, analysed all the
relevant facts and solved the problem completely in 1939.
A year earlier,Weisszacker had given a partial solution.

Bethe's paper is a masterpiece \cite{2}. It gave a  complete
picture of the thermonuclear reactions that power the
Sun and the stars. However, a not-so-well-known fact
is that Bethe leaves out the neutrino that is emitted
along with the electron, in the reactions enumerated
by him. Neutrino, born in Pauli's mind in 1932, named
and made the basis of weak interaction by Fermi in
1934, was already a well-known entity in nuclear
physics. And it is Fermi's theory that Bethe used in
his work. So it is rather inexplicable why he ignored
the neutrinos in his famous paper. The authority of
Bethe's paper was so great that the astronomers and
astrophysicists who followed him in the subsequent
years failed to note the presence of neutrinos. Even
many textbooks in Astronomy and Astrophysics written
in the 40's and 50's do not mention neutrinos! This
was unfortunate, since we must realize that, in spite
of the great success of Bethe's theory, it is
nevertheless only a theory. Observation of neutrinos
from the Sun is the only direct experimental
evidence for Eddington's thermonuclear hypothesis
and Bethe's theory of energy production. That is
the importance of detecting solar neutrinos.

The basic process of thermonuclear fusion in the Sun and stars
is four protons (which are the same as Hydrogen nuclei)
combining into a Helium nucleus and releasing
two positrons, two neutrinos and 26.7 MeV of energy.
\begin{center}
$ p + p + p + p  \rightarrow  He^{4} + e^{+} + e^{+} + \nu_{e} + \nu_{e}$
\end{center}

This can be regarded as the most important reaction for all life,
for without it Sun cannot shine and there can be no life on Earth! 

However,the probability of four protons meeting at a point
is negligibly small even at the large densities existing
in the solar core. Hence the actual series of nuclear reactions
occurring in the solar and stellar cores are given by the
so-called carbon cycle and the pp-chain. In the carbon cycle
the four protons are successively absorbed in a series of
nuclei,starting and ending with carbon. In the pp-chain two
protons combine to form the deuteron and further protons
are added.

We shall not go into details here \cite{3} except noting that
both in the carbon cycle and the pp-chain, the net process is
the same as what was mentioned above, namely the fusion of
four protons to form alpha particle with the emission of two
positrons and two neutrinos.

It is these thermonuclear fusion reactions that are responsible 
for the Sun and the stars continuing to shine for billions
of years. This fact remained as a theoretical fact for many
decades although it was accepted as generally correct by 
scientists. So even Nobel Prize was given to Bethe in 1967.

The only way to prove Bethe's theory is to detect the neutrinos
coming from the Sun.

It is easy to calculate from the solar luminosity the total
number of neutrinos emitted by the Sun; for `every' 26.7 MeV
of energy received by us, we must get 2 neutrinos. Thus one gets 
the solar neutrino flux at the earth as 70 billion per square
cm per sec. These many solar neutrinos are passing through our body
and the Earth.

\section{The Davis Experiment}

About 50 years ago, Ray Davis started his pioneering experiments
to detect the solar neutrinos. His experiment was 
based on the inverse beta decay:
\begin{center}
$\nu_{e} + Cl^{37}  \rightarrow  e^{-} + Ar^{37} $
\end{center}

Chlorine-37 absorbs the solar neutrino to yield Argon-37 and
an electron. (See Section 4 for explanation of beta 
decay and inverse beta decay.)

A tank containing 615 tons of a fluid rich in chlorine
called tetrachloroethylene was placed in the Homestake gold mine
in South Dakota(USA). The Chlorine-37 atoms in the fluid were
converted into Argon-37 atoms by the above reaction.
The fluid was periodically purged with
Helium gas to remove the Argon-37 atoms which were then counted
by means of their radioactivity. Davis continued his experiment
for almost 30 years and the result was that about one neutrino
in three days was caught in his experiment. 

Two points must be noted. In three days billions of neutrinos
fall on Davis's tank, but only one among them reacted with 
Chlorine-37 and got caught. All others escape without any interaction, 
thus showing how tiny is the probability of interaction of a neutrino. 
The experiment also proves the extraordinary capability of 
Davis in counting radioactive atoms. If you colour one grain of
sand red and mix it in the sand of Sahara desert, can one
find that red grain of sand? The achievement of Davis is
comparable to that.

Although solar neutrinos were detected by Davis, a new puzzle
appeared. Actually Davis detected only about a third of the
solar neutrinos that must have been detected in his tank.
What is the reason for this discrepancy between the theoretical
number of solar neutrinos that must be detected in Davis's
detector and the actual number detected? Are the thermonuclear
fusion hypothesis and Bethe's theory based on it wrong? This
became known as the solar neutrino puzzle and the puzzle lasted 
for many years.

\section{Kamioka and Superkamioka}

A few other experiments were undertaken in the attempt to
resolve the solar neutrino puzzle. The most important one
among them was the Kamioka experiment in Japan led by 
Matoshi Koshiba. 

One must also note that
Davis's radio chemical experiment was a passive experiment.There
was actually no proof that he detected any solar neutrinos.In
particular if a critic claimed that all the radioactive atoms
that he detected were produced by some background radiation,
there was no way of conclusively refuting it. That became
possible through the Kamioka experiment that went into operation
in the 80's. 

In contrast to Davis's chlorine tank which was a passive
detector,the Kamioka water Cerenkov detector is an active real time detector. Solar neutrino kicks out an
electron in the water molecule by elastic scattering and the
electron is detected through the Cerenkov radiation it emits.
Since the electron is mostly kicked toward the forward direction,
the detector is directional.A plot of the number of events against
the angle between the electron track and Sun's direction gives
an unmistakable peak at zero angle,proving that neutrinos from
the Sun were being detected.The original Kamioka detector had
2 kilotons of water and the Cerenkov light was collected by an
array of 1000 photomultiplier tubes, each 20" diameter and this
was later superceded by the SuperKamioka detector which had
50 kilotons of water faced by 11,000 photomutiplier tubes.
Both Kamioka and SuperK gave convincing proof of the detection
of solar neutrinos.The ratio of the measured solar
neutrino flux to the predicted flux was about 0.5, thus
confirming the solar neutrino puzzle.

There is a difficulty in resolving the solar neutrino puzzle.
To understand that, we have to know more details about the Sun.

\section{Standard Solar Model and the Gallium Experiment}

In the Sun, the dominant thermonuclear fusion process is the 
pp-chain.  Although the 70 billion neutrinos per square centimeter per sec as the total number of solar neutrinos falling on the Earth could be trivially
calculated from the solar luminosity,their energy spectrum which
is crucial for their experimental detection,requires a detailed
model of the Sun, the so-called Standard Solar Model (SSM).
SSM is based on the thermonuclear hypothesis and Bethe's theory,
but uses a lot more physics input about the interior of the Sun.

A knowledge of the neutrino energy spectrum is needed since
the neutrino detectors are strongly energy sensitive.In fact all
detectors have an energy threshold and hence miss out the very
low energy neutrinos.

Leaving out the details \cite{3}, the solar neutrino spectrum is
roughly characterized by a dominant (0.9975 of all neutrinos)
low energy spectrum ranging from 0 to 0.42 MeV and a very weak
(0.0001 of all the neutrinos)
high energy part extending from 0 to 14 MeV. Most of the
neutrino detectors detect only the tiny high-energy branch of
the spectrum.

While the dominant low-energy neutrino flux is basically
determined by the solar luminosity,the flux of the high-energy
neutrino flux is very sensitive to the various physical
processes in the Sun and hence is a test of SSM. In fact,this
latter flux is a very sensitive function of the temperature
of the solar core,being proportional to the 18th power of
this temperature and hence this neutrino flux provides
a very good thermometer for the solar core. In contrast to
the photons which hardly emerge from the core,the neutrinos
escape unscathed and hence give us direct knowledge about
the core.

There is a simple physical reason for this sharp dependence
on temperature. It is related to the quantum-mechanical
tunnelling formula, the famous discovery of George Gamow.
The probability for tunnelling through the repulsive Coulomb
barrier has a sharp exponential dependence on the kinetic
energy of the colliding charged particles.

The detection threshold in Davis's experiment was 0.8 MeV and
thus only the high-energy neutrinos were detected. SSM
could be used to get the number of neutrinos expected above
this threshold and the detected number was less than the
predicted number by a factor of about 3. Over the three decades
of operation of Davis's experiment,this discrepancy has
remained and has been known as the solar neutrino puzzle.

The energy threshold of the Kamioka and SuperKamioka detectors
was about 7 MeV and so only the high-energy part of the neutrino
spectrum was being detected.

The next input came from the gallium experiments. The high-energy
neutrino flux is very sensitive to the details of the SSM and
so SSM could be blamed for the detection of a lower flux.On the
other hand the low energy neutrinos are not so sensitive to
SSM. So the gallium detector based on the inverse beta decay of
Gallium-71 was constructed.Although this was also a passive
radiochemical detector,its threshold was 0.233 MeV and hence
it was sensitive to a large part of the low-energy branch extending up to
0.42 MeV.Actually two gallium detectors were mounted,called
SAGE and GALLEX and both succeeded in detecting the low energy neutrinos
in addition to the high energy neutrinos but again at a depleted level
by a factor of about 0.5.

To sum up, there were three classes of neutrino detectors with
different energy thresholds,all of which detected solar neutrinos,
but at a depleted rate.The ratio R of the measured flux to the
predicted flux was 0.33$\pm$0.028 in the chlorine experiment,
0.56$\pm$0.04 in the two gallium experiments (average) and
0.475$\pm$0.015 in the SuperK experiment.

Actually it must be regarded as a great achievement for both
theory and experiment that the observed flux was so close to the
theoretical one, especially considering the tremendous amount of
physics input that goes into the SSM. After all R does not differ
from unity by orders of magnitude! This is all the more significant
since the large uncertainties in some of the low energy thermonuclear
cross sections do lead to a large uncertainty in the SSM prediction.
But astrophysicists led by John Bahcall were ambitious and claimed
that the discrepancy is real and must be explained. Two points
favour this view.As already stated,the gallium experiments
sensitive to the low-energy flux which is comparatively free of the
uncertainties of SSM, also showed a depletion in the flux. Second,
SSM has been found to be very successful in accounting for many
other observed features of the Sun, in particular the 
helioseismological data i.e data on solar quakes.

Hence something else is the reason for R being less than unity
and that is neutrino oscillation.

\section{Three kinds of Neutrinos}

In addition to the well-known electron,two heavier types of
electrons are known to exist.Reserving the name electron to
the well-known particle of mass 0.5 MeV,the heavier ones are
called muon ($\mu$) and tauon ($\tau$) and their masses are 105 and 1777 MeV
respectively. Correspondingly there are three types or flavours
of neutrinos called eneutrino ($\nu_{e}$), mu neutrino ($\nu_{\mu}$) 
or tau neutrino ($\nu_{\tau}$) that are respective companions of
electron,muon or tauon. Just is electron and eneutrino are
emitted in beta decay, in the processes involving muon or taon,
muneutrino or tauneutrino will appear.  The three doublets are shown below:
\[ \left( \begin{array}{c}
\nu_e  \\
e   \end{array} \right) 
\left( \begin{array}{c}
\nu_\mu  \\
\mu   \end{array} \right)
\left( \begin{array}{c}
\nu_\tau  \\
\tau   \end{array} \right)
\]

What is produced in the thermonuclear reactions in the Sun
is the antielectron (positron) and eneutrino. This eneutrino
produced an electron when it converted the Chlorine-37 nucleus 
in Davis's detector into an Argon-37 nucleus:
\begin{center}
$\nu_{e} + Cl^{37}  \rightarrow  e^{-} + Ar^{37} $	
\end{center}

If some of the eneutrinos oscillate
to the muneutrinos or the tauneutrinos on the way to the earth,
the reactions in Davis's detector must be
\begin{center}
$\nu_{\mu} + Cl^{37}  \rightarrow  \mu^{-} + Ar^{37} $\\
or\\
$\nu_{\tau} + Cl^{37}  \rightarrow  \tau^{-} + Ar^{37} $
\end{center}
Just as the eneutrino
produces an electron in the inverse beta decay process,
the muneutrino or the tauneutrino has to produce a muon or a tauon
respectively in the final state. But since the energy of the
solar neutrinos are limited to 14 MeV,the muon or tauon with
the high masses of 105 and 1777 MeV cannot be produced in
the inverse beta decay. According to Einstein's famous equation,
$$ E = m c^{2} $$
\noindent
it is energy E which is converted into mass m. 
So the neutrinos that have been converted into the mu or tau flavour through oscillation escape detection in the Chlorine and Gallium experiments.

Although elastic scattering of neutrinos on electron which
is used as the detecting mechanism in the Kamioka and SuperK
water Cerenkov detectors can detect the converted mu or tau
flavours also, it has a much reduced efficiency. Hence the
depletion of the number of neutrinos observed in the water
detector also is attributable to oscillation.

There was a famous painting called "The Cow and Grass".But
nothing except a blank convass was visible.When asked to
show the grass, the painter said the cow had eaten the grass.
When pressed to show at least the cow,he said it went away
after eating the grass.

Our neutrino story so far is like that.We said thermonuclear reactions
in the Sun must produce so many neutrinos.We did not see so
many neutrinos, but then explained them away through oscillations.

In Science we have to do something better.If we say that neutrinos have
oscillated into some other flavour, we have to see the
neutrinos of those flavours too.

This is precisely what is done in a two-in-one experiment.

\section{Two-in-One Experiment}

The beta decay and inverse beta decay processes that we have described 
so far are charge-changing (CC) weak interaction processes. Another
kind of weak interaction, known as charge-nonchanging or neutral
current (NC) weak interaction was discovered in 1973. These two
kinds of processes are shown below:
\begin{center}
$ \nu_{e} + (Z,N) \rightarrow  (Z+1, N-1) + e^{-} $      (CC) \\
$ \nu + (Z,N) \rightarrow (Z,N)^{*} +\nu  $             (NC) 
\end{center}

In the CC process eneutrino changes into electron. Neutrino
does not have charge while electron does have charge. So charge
of the particle changes in the process and hence CC. The nucleus
also changes from (Z,N) to (Z+1, N-1) and so its charge changes.
But in the NC process, neutrino remains as neutrino. The
nucleus (Z,N), without changing its charge, either gets exited
to a higher energy state or disintegrates.
We have denoted such a state of the nucleus as $(Z,N)^{*}$ in the
NC reaction above.  

The important point is that the solar eneutrinos that oscillated 
into the mu type or the tao type cannot undergo
the appropriate CC process as we already explained. But since 
the NC process does not create the heavier muon or taon, they
can undergo the NC process.
So if we design an experiment in which both the CC and NC modes
are detected, and if the number of neutrinos involved in NC 
reactions is found to be larger than those in CC reactions, 
oscillation will be proved. 

While the CC mode will give the number
of eneutrinos, the NC mode will give the total number of e, mu and
tau type of neutrinos. 
The total number detected will be a test of
SSM independent of oscillations while the NC minus CC events
will give the number that had oscillated away.

This is the 'two-in-one" experiment.
A huge two-in-one detector based on Boron called BOREX was 
proposed by Sandip Pakvasa and  Raju Raghavan (who passed away 
in 2011), but that has not materialized.The two-in-one detector
based on deuteron in heavy water proposed by Chen was constructed
at the Sudbury Neutrino Observatory (SNO), Canada and it finally
solved the solar neutrino puzzle.
SNO uses 1000 tons of heavy water borrowed from the Canadian
Atomic Energy Commission. 

Just as water is made of $H_{2}O$ molecules, heavy water is made
of $ D_{2}O $ molecules. The nucleus of the heavy hydrogen D is made
up of one proton and one neutron.
Solar neutrino breaks up the
deuteron D by CC and NC modes. While CC mode leads to two protons
and an electron, NC mode leads to a neutron, a proton and a neutrino.
\begin{center}
$ \nu_{e} + D \rightarrow p + p + e^{-} $    (CC) \\
$ \nu  + D  \rightarrow  p + n + \nu $     (NC) \\
\end{center}

The threshold of detection was again high like SuperK so that 
only the high energy neutrinos were detected.
Let us now straightaway go to the exciting results
of SNO that came out in April 2002.

The CC mode gave the flux (million neutrinos per sq cm per sec)
as 1.76$\pm$0.11 while the NC gave 5.09$\pm$0.65 in the same units.
(The numbers are in millions rather than in billions since
the threshold of detection was again high like in SuperK so that
only the high-energy neutrinos were detected.)
Thus we conclude that the flux of e + mu + tau neutrinos is
5.09$\pm$0.65 while that of the e flavour alone is 1.76$\pm$0.11.
The difference 3.33$\pm$0.66 is the flux of the mu + tau flavours.
Hence oscillation is confirmed. Roughly two third of the 
eneutrinos have oscillated to the other flavours. Further,
comparing with the SSM prediction of 5.05$\pm$0.40, SSM also
is confirmed. So at one sweep the SNO results
confirmed both the SSM based on the thermonuclear fusion hypothesis
and neutrino oscillation.

What is the moral of the story? When we said in the beginning
that the thermonuclear hypothesis for the Sun has to be proved,
it was not a question of proof before a court of law. Science
does not progress that way. In trying to prove the hypothesis
experimentally through the detection of solar neutrinos, Davis
and the other scientists have made a discovery of
fundamental importance, namely that the neutrinos have mass.
Only if they have mass, they can oscillate.

\section{Neutrino Oscillation}

To understand neutrino oscillation, one must think
of neutrino as a wave rather than a particle
(remember quantum mechanics).Neutrino oscillation
is a simple consequence of its wave property.Let
us consider the analogy with light wave.Consider a
light wave travelling in the z-direction. Its 
polarization could be in the x-direction, y-direction
or any direction in the x-y plane.This is the case
of plane-polarized wave. However the wave could have
circular polarization too, either left or right.
Circular polarization can be composed as a linear
superposition of the two plane polarizations in the
x and y directions. Similarly plane polarization can
be regarded as a superposition of the left and right
circular polarizations.

Now consider plane polarized wave travelling through an optical medium.
During propagation through the medium, it is important to resolve the
plane polarized light into its circularly polarized components
since it is the circularly polarized wave that has well-defined
propagation characteristics such as the refractive index or velocity of
propagation. In fact in an optical medium waves with the left
and right circular polarizations travel with different velocities.
And so when light emerges from the medium, the left and right
circular polarizations have a phase difference proportional
to the distance travelled. If we recombine the circular
components to form plane polarized light, we will find the
plane of polarization to have rotated from its initial
orientation. Or, if we start with a polarization in the
x-direction, a component in the y-direction would be
generated at the end of propagation through the optical medium.

For the neutrino wave, the analogues of the two planes of
polarizations of the light wave are the three flavours (e,
mu or tau) of the neutrino (see table~\ref{tab:table1}). When the neutrinos are produced
in the thermonuclear reactions in the solar core, they are
produced as the e type. When the neutrino wave propagates,
it has to be resolved into the analogues of circular polarization
which are energy eigenstates or mass eigenstates of the neutrino.
These states have well-defined propagation characteristics with
well-defined frequencies (remember frequency is the same as energy
divided by Planck's constant). The e type of neutrino wave will
propagate as a superposition of three mass eigenstates which
pick up different phases as they travel. At the detector, we
recombine these waves to form the flavour states.Because of
the phase differences introduced during propagation, the
recombined wave will have rotated "in flavour space". In general,
it will have a mu component and tau component in addition to
the e component it started with. This is what is called neutrino
oscillation or neutrino flavour conversion through oscillation.

\begin{table}[h!]
	\renewcommand{\arraystretch}{1.2}
	\centering
	\setlength{\tabcolsep}{20pt}
	\begin{tabular}[c]{ c  c}
		\toprule
		Light wave & Neutrino wave \\
		\toprule
		
		Plane polarization & Flavour state \\
		x or y             & $\nu_e$, $\nu_\mu$ or $\nu_\tau $ \\
		\midrule
		Circular polarization & Mass eigenstate \\
		right or left & $\nu_1$, $\nu_2$ or $\nu_3$ \\
		\bottomrule
	\end{tabular}
		\caption{The analogy between light wave and neutrino wave}
		\label{tab:table1}
\end{table}

Flavour conversion is directly due to the phase difference
arising from the frequency difference or energy difference
which in turn is due to the mass difference. Mass difference
cannot come without mass.Hence discovery of flavour conversion
through neutrino oscillation amounts to the discovery of
neutrino mass. This is the fundamental importance of neutrino
oscillation, since so far neutrinos were thought to be massless
particles like photons.

Since it is an oscillatory phenomenon, the probability of
flavour conversion is given by oscillatory functions of the
distance travelled by the neutrino wave, the characteristic
"oscillation length" being proportional to the average energy
of the neutrino and inversely proportional to the difference
of squares of masses. Further,the overall probability for 
conversion is controlled by the mixing coefficients that
occur in the superposition of the mass eigenstates to form
the flavour states and vice versa. These mixing coefficients
form a 3x3 unitary matrix.

Neutrino oscillations during neutrino propagation in matter
become much more complex and richer in physics, but we shall
not go into the details here. After Wolfenstein calculated
the important effect of matter on the propagating
neutrino and Mikheyev and Smirnov drew attention to the dramatic
effect on neutrino oscillation when the neutrino passes through
matter of varying  density, it was Bethe who gave an elegant
explanation of the MSW (Mikheyev-Smirnov-Wolfenstein) effect
based on quantum mechanical level-crossing. In fact most people
(including the present author) appreciated the beauty of MSW effect
only after Bethe's paper came out. One may comment that Bethe
redeemed himself for his earlier omission of neutrinos in his famous paper
on the energy production in stars.

We next go to the cosmic-ray-produced neutrinos since their study
and its interplay with solar neutrino research constitute a fascinating
chapter in the story of the neutrino.

\section{Cosmic-Ray-Produced or Atmospheric neutrinos}

\begin{figure}[h!]
	\centering
%	\label{fig1}
	\includegraphics[scale=0.4]{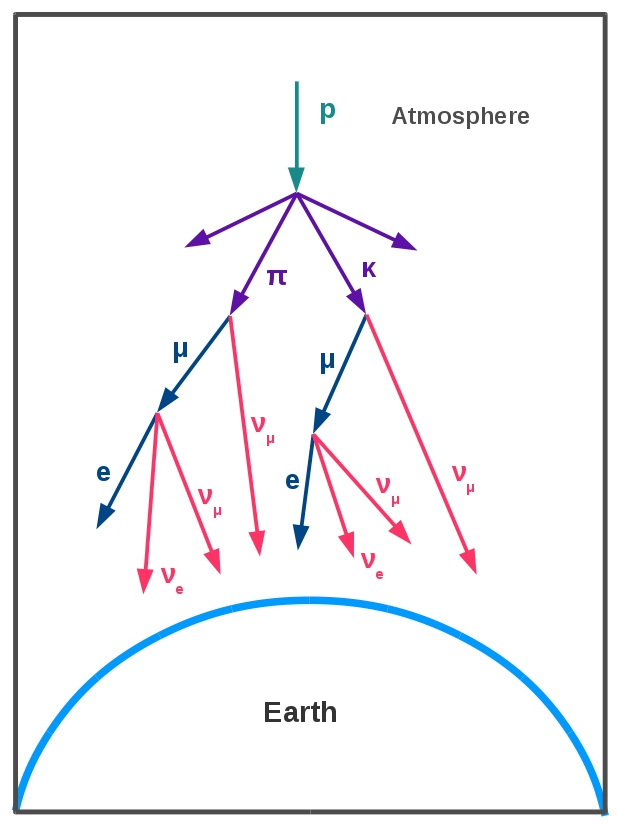}
	\caption{Cosmic Ray Shower}
	\label{fig1}
\end{figure}

Cosmic Rays were discovered around the year 1900. They are mostly 
very energetic protons. They are created in many parts of the Universe
and are flying in all directions everywhere. They fall on Earth too.
Since Earth is surrounded by atmosphere, these protons collide on the
nitrogen or oxygen nuclei of the atmosphere and in these collisions
many kinds of elementary particles are created. All these move
in the direction of the Earth. Figure~\ref{fig1} \ shows such a cosmic-ray shower.
Many elementary particles such as muon ($\mu $), pion ($ \pi $), and Kaon (K)
were originally discovered in cosmic ray research only. As seen from 
the Figure~\ref{fig1}, all these particles decay and give rise to neutrinos.
They are cosmic-ray produced neutrinos although they are generally
called atmospheric neutrinos.

Homi Jahangir Bhabha who founded the Tata Institute of Fundamental
Research in Mumbai was well-known for cosmic ray research. Around
1950, he suggested to B V Sreekantan that cosmic ray research must 
be conducted in Kolar Gold Field (KGF) mine which is one of the 
deepest mines in the world. His idea was to measure the flux of cosmic ray
particles as we go down the depth of one or two kilometers below
the Earth and verify experimentally whether the penetrating component of the
cosmic rays was composed of muons alone (as he had concluded in his
earlier theoretical research) or whether there was any other particle.

Sreekantan, Ramanamurthy and Naranan followed Bhabha's suggestion and
thus started the pioneering KGF experiments and the experiments continued
for more than two decades. The scientists determined how the muon flux
decreased as a function of the depth. When the experiments were continued
at greater and greater depths, at a certain depth the number of the muons
detected became zero. At that depth (which was about 2 kilometer from
the surface of the Earth) all the muons are absorbed by the rock above,
but neutrinos are not absorbed and hence could be detected without any
disturbance from other particles such as muons. The scientists succeeded
in detecting these neutrinos. This happened in the year 1965. This was the
first detection of cosmic-ray-produced neutrinos in the world. The credit
for this achievement goes to the Tata Institute of Fundamental Research
and two other collaborating institutions Durham University, UK and
Osaka University, Japan. Last year 2015 was the Golden Jubilee
Year of this milestone in the story of the neutrino.

The atmospheric neutrino research that started in India progressed further
especially in Japan and brought great success to the Japanese physicists.
We have already described how the Kamioka and SuperKamioka experiments
succeeded in catching the solar neutrinos. The same experiments caught
the atmospheric neutrinos also. Further study led to another
discovery which we describe now.

The pion born from cosmic rays decays into a muon and a muneutrino.
Then the muon also decays into an electron, an eneutrino and a 
muneutrino. The decay of the Kaon also leads to same results. Hence
as shown in Figure~\ref{fig1}, the number of muneutrinos reaching the Earth is
twice the number of eneutrinos. In the Kamioka experiments, it
was possible to distinguish the two kinds of neutrinos. Since the
cosmic ray protons had a very high energy, about 1000 MeV, the neutrinos
born from them have very high energy and so can create the muons of
107 MeV. The muneutrinos colliding with the nuclei in the detector
produce muons and eneutrinos produce electrons. Since 
muons and electrons emit different kinds of Cerenkov light, the
Kamioka and SuperK experimenters succeeded in counting the number
of colliding muneutrinos and eneutrinos separately.

\begin{figure}[h!]
	\centering
%	\label{fig2}
	\includegraphics[scale=0.5]{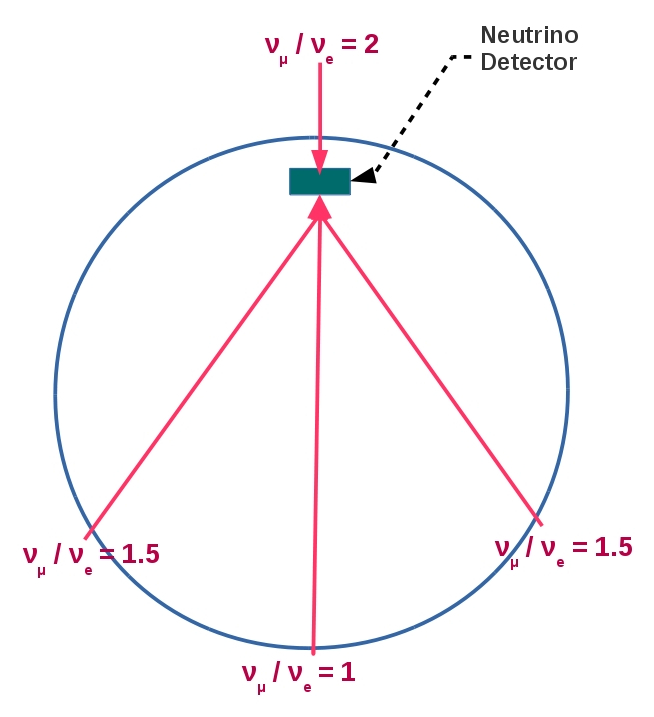}
	\caption{Atmospheric Neutrinos at the SuperK Detector}
	\label{fig2}
\end{figure}

The underground SuperK detector and the directions in which neutrinos
arrive at the detector are shown in Figure~\ref{fig2}. The sky and atmosphere
surround the Earth in all directions and so the neutrinos arrive from
all directions. In the downward direction, the ratio of muneutrinos
to eneutrinos was experimentally shown to be 2 as expected. But this
ratio gradually decreased from 2 as the direction changed and became 
unity for the upward moving neutrinos. Although Kamioka detector
and a few other detectors saw this anomaly in 1990, it required
the bigger SuperK detector with its superior statistics to establish
the effect in 1998.

About half of the upward moving muneutrinos have disappeared. How?
The maximum height of the atmosphere is about 20 kilometer. So 
neutrinos coming downwards from above travel only a few kilometers
and reach the detector without oscillation. Neutrinos coming upwards
have to cross a distance of 13,000 kilometers which is Earth's diameter and 
undergo oscillation. Half of the muneutrinos oscillate to the
tauneutrinos. Although the cosmic-ray produced neutrinos have
high enough energy to create the muon, their energy is not
sufficient to create the taon of mass 1777 MeV. So the tauneutrinos
escape undetected. Thus the SuperK experiment discovered the
oscillation of cosmic-ray produced neutrinos.

\section{The Nobel Prizes: Solar and Atmospheric Neutrinos}

One may say that it is in the Davis experiment on solar neutrinos
that neutrino oscillation and neutrino mass were discovered first.
However it was not possible to accept these conclusions as firm
on the basis of the Davis experiment. For, as we mentioned earlier
the question as to whether the flux of the higher energy neutrinos
from the Sun was calculated correctly could not be settled without
any doubt. This doubt was completely removed only by the results
of the two-in-one experiment of SNO, since the inference of neutrino
oscillation from SNO results was completely independent of the
calculation of the solar neutrino flux.

SNO results came out only in 2002. Much before that, in 1998,
SuperK discovered the oscillations of cosmic-ray-produced
neutrinos. Their discovery concerned the ratio of muneutrinos
to eneutrinos and hence did not depend on the uncertainties of
calculated fluxes of the neutrinos produced by cosmic rays. Hence
it was accepted that the discovery of neutrino oscillation and 
neutrino mass by SuperK in cosmic-ray-produced neutrino experiments
was free from doubts of the kind that plagued the interpretation
of Davis and SuperK experiments on solar neutrinos. In the race 
for the discovery of oscillations experiments on cosmic-ray-produced 
neutrinos won over those on solar neutrinos.

In 2002, Nobel Prize was given to Ray Davis who pioneered solar
neutrino research, was the first to detect solar neutrinos and 
continued the experiments for more than 30 years and Matoshi Koshiba
who was the leader of the Kamioka and SuperK experiments that 
detected solar neutrinos, cosmic-ray-produced neutrinos and
Supernova neutrinos. The Nobel Prize of 2015 was given to Arthur
McDonald who was the leader of SNO which proved thermonuclear
fusion as the source of solar energy and firmly established
oscillation of solar neutrinos and to Takaaki Kajita who was the
leader of SuperK that discovered the oscillations of cosmic-ray-
produced neutrinos.

\section{Neutrino masses and Mixing}

As we already mentioned, nuclear reactors produce antineutrinos
copiously. High energy protons from particle accelerators
produce pions whose decays ultimately lead to neutrinos. This
is in fact the same process as in case of cosmic-ray protons which we
mentioned earlier. 

Solar neutrinos, atmospheric neutrinos, reactor neutrinos
and accelerator neutrinos -- many experiments on all these have
been done and considerable amount of information on neutrino
oscillations have been learnt. Most importantly, the mass-differences
between the three kinds of neutrinos have been determined and they
are very very tiny:
\begin{center}
$ m^{2}_{2} - m^{2}_{1} = 0.00007 eV^{2} $ \\
$ | m^{2}_{3} - m^{2}_{2} | =0.002 eV^{2} $ \\ 
\end{center}

Note one of the mass difference is known only in magnitude and its
sign has yet to be determined and so the ordering of the three mass
levels is not yet known.

Oscillation experiments give only mass differences. To determine
the mass itself a different kind of experiment has to be done.
From the precision experimental study of the continuous energy
distribution of the electrons emitted in the beta decay of 
Tritium (heavy Hydrogen), an upper limit of 2.2 eV for the neutrino
mass or masses has been determined. So, all the three neutrino
masses are clustered close to each other at a value smaller than
2.2 eV. Among all the massive elementary particles, electron
has the lowest mass 0.5 MeV. Neutrino masses are a million times
smaller. But many secrets of the Universe are hidden in this tiny
number.

As a culmination of hundred years of fundamental research
a theory called the Standard Model of High Energy Physics \cite{4}
has been shown to be the basis of almost All of physics except
gravity. But according to this theory neutrinos are massless.
Hence the importance of the discovery that neutrinos have mass.
Neutrino mass may be the portal to go beyond Standard Model. 

The oscillation experiments also determined the 3x3 mixing
matrix that tells us how the three massive neutrinos are 
superposed to give the three flavours e, $\mu$, $\tau$
of neutrinos. This unitary matrix is characterized by
three angle parameters and a phase. The values of the three 
angles as determined by the oscillation experiments \cite{5} are
\begin{center}
$ \theta_{12} = 30^\circ $ \\
$ \theta_{23} = 45^\circ $ \\
$ \theta_{31} = 9^\circ \ $ \\
\end{center}

The phase however is not yet determined. This phase is very 
important since it signals matter-antimatter asymmetry which
in turn can play an important role in the evolution of the Universe
as pointed out below.

\section{Continuing Story}

There are many more things in the neutrino story. We shall
describe them briefly.

Generally there is an antiparticle for every particle. This
is a Law of Nature which is a consequence of combining
quantum mechanics with relativity and was discovered by 
Dirac. Positron is the antiparticle of electron. Their
electric charges are equal in magnitude but opposite in sign.
However, when the electric charge is zero as is the case for
neutrino, its antiparticle, namely the antineutrino could
be the same as the neutrino itself. If this is true, the
particle is called a Majorana particle, named after Majorana
who envisaged such a possibility. A particle whose antiparticle
is different, such as the electron is called a Dirac particle.
Is neutrino a Majorana particle?\cite{6} This is the most important
question in Neutrino Physics and this question can be answered only
by the `neutrinoless double beta decay experiment'. These experiments
are going on, but have not yet yielded a definitive answer. 

Cosmologists have found good evidence that the Universe was
born 14 billion years ago in a gigantic explosion called 
the Big Bang. At that point, the Universe must have contained
equal number of particles and antiparticles. However there
are only particles now. All the atoms in the Universe are
made of protons, neutrons and electrons only. What happened
to the antiprotons, antineutrons and positrons? How did they
disappear? How was the matter-antimatter symmetry that existed
at the beginning of the Universe destroyed? This is an
important cosmological puzzle. The key to solving this puzzle
is contained in the neutrino. If neutrino and antineutrino can be proved
to be the same and if the phase in the mixing matrix (see above) is proved
to be nonzero, this puzzle can be answered. Hence, neutrino plays
an important role in cosmological research.   

Supernova explosion is the end stage of most of the stars.
Most of the energy of the explosion is released through the
neutrinos that are emitted in a very large number. The
neutrinos emitted in the so-called Supernova 1987-a were
detected in the SuperK detector. This was one of the reasons
for the Nobel Prize given to Koshiba in 2002 since this was the first time
neutrinos from outside the solar system were detected on the Earth
and supernova neutrino research was thus initiated. 

Ultrahigh energy neutrinos with energy greater than $10^{12}$ eV
coming from outer space have been detected in the year 2013. This was achieved
by using ice as a detector in the Antarctica Continent near the South Pole.
The size of this ice detector is one kilometer in length, one kilometer
in breadth and one kilometer in height and it is called Ice Cube.   

Radioactive Uranium and Thorium ores lying buried in the deep bowels of the
Earth emit neutrinos. These geoneutrinos have been detected in the 
KamLAND detector in Japan and the BOREXINO detector of the 
Gran Sasso laboratory in Italy. Through this, one can map 
where and at what depths Uranium and Thorium ores lie and this
knowledge will be used in Geochronology. Thus a new window on
Earth Science has been opened by neutrino research

The bulk of the low-energy neutrinos constituting more than 90 percent
of the solar neutrinos which had eluded detection have now been detected 
by the BOREXINO detector. The measured flux is in very good agreement
with SSM.

Neutrinos are the most penetrating radiation known to us. A typical
neutrino can travel through a million earth diameters without getting stopped.
However because of the MSW effect the neutrino senses the density profile
of the matter through which it travels and so the flavour composition
of the final neutrino beam can be decoded to give information about the
matter through which it has travelled. Hence tomography of the Earth's interior
through neutrinos will be possible which may even lead to the prediction
of earthquakes in future. This requires our mastery of neutrino technology.
But neutrino technology will be mastered and neutrino tomography will come.

Efforts are going on all over the world to create new underground
laboratories for neutrinos. As already mentioned, India was a
pioneer in neutrino research. The cosmic-ray-produced neutrinos
first detected in KGF in India in 1965 led to two Nobel Prizes for the Japanese
physicists. But the KGF mines were closed in 1995. To recover this lost
initiative the India-based Neutrino Observatory (INO) has been planned [7].
The underground laboratory will be created in a huge cavern to be dug out in
a mountain in Theni District near Madurai and the main Centre of INO will be 
built in Madurai City. In the first stage, a neutrino oscillation experiment
using atmospheric neutrinos will be performed in a gigantic 50,000 ton magnetised
iron detector which will be mounted inside the underground laboratory \cite{7}.

\section*{Milestones in the neutrino story}
\addcontentsline{toc}{section}{Milestones in the neutrino story}
\noindent
1930 Birth of Neutrino: Pauli \\
1932 Theory of beta decay, "Neutrino" named: Fermi \\
1954 First detection of neutrino: Cowan and Reines \\
1964 Discovery of muneutrino: Lederman, Schwartz and Steinberger \\
1965 Detection of atmospheric neutrino: KGF \\
1970 Start of the solar neutrino experiment: Davis \\
1987 Detection of neutrinos from supernova: SuperKamioka \\  
1998 Discovery of neutrino oscillation and mass: SuperKamioka \\
2001 Discovery of tauneutrino: DONUT \\
2002 Solution of the solar neutrino puzzle: SNO \\
2005 Detection of geoneutrinos: KamLAND \\
2013 Detection of ultra high energy neutrinos from space: Ice Cube \\

\end{document}